\documentclass[12pt]{article}
\usepackage{graphicx}
\usepackage{color}

\def\hybrid{\topmargin 0pt      \oddsidemargin 0pt
        \headheight 0pt \headsep 0pt
       \voffset-1cm
        \textwidth 6.25in       
       \textheight 9.5in       
        \marginparwidth 0.0in
        \parskip 5pt plus 1pt   \jot = 1.5ex}
\catcode`\@=11
\def\marginnote#1{}

\newcount\hour
\newcount\minute
\newtoks\amorpm
\hour=\time\divide\hour by60
\minute=\time{\multiply\hour by60 \global\advance\minute by-\hour}
\edef\standardtime{{\ifnum\hour<12 \global\amorpm={am}%
        \else\global\amorpm={pm}\advance\hour by-12 \fi
        \ifnum\hour=0 \hour=12 \fi
        \number\hour:\ifnum\minute<10 0\fi\number\minute\the\amorpm}}
\edef\militarytime{\number\hour:\ifnum\minute<10 0\fi\number\minute}

\def\draftlabel#1{{\@bsphack\if@filesw {\let\thepage\relax
   \xdef\@gtempa{\write\@auxout{\string
      \newlabel{#1}{{\@currentlabel}{\thepage}}}}}\@gtempa
   \if@nobreak \ifvmode\nobreak\fi\fi\fi\@esphack}
        \gdef\@eqnlabel{#1}}
\def\@eqnlabel{}
\def\@vacuum{}
\def\draftmarginnote#1{\marginpar{\raggedright\scriptsize\tt#1}}

\def\draftlabel#1{{\@bsphack\if@filesw {\let\thepage\relax
   \xdef\@gtempa{\write\@auxout{\string
      \newlabel{#1}{{\@currentlabel}{\thepage}}}}}\@gtempa
   \if@nobreak \ifvmode\nobreak\fi\fi\fi\@esphack}
        \gdef\@eqnlabel{#1}}
\def\@eqnlabel{}
\def\@vacuum{}
\def\draftmarginnote#1{\marginpar{\raggedright\scriptsize\tt#1}}

\def\draft{\oddsidemargin -.5truein
        \def\@oddfoot{\sl preliminary draft \hfil
        \rm\thepage\hfil\sl\today\quad\militarytime}
        \let\@evenfoot\@oddfoot \overfullrule 3pt
        \let\label=\draftlabel
        \let\marginnote=\draftmarginnote
   \def\@eqnnum{(\theequation)\rlap{\kern\marginparsep\tt\@eqnlabel}%
\global\let\@eqnlabel\@vacuum}  }

\def\numberbysection{\@addtoreset{equation}{section}
        \def\theequation{\thesection.\arabic{equation}}}

\def\underline#1{\relax\ifmmode\@@underline#1\else
        $\@@underline{\hbox{#1}}$\relax\fi}

\def\titlepage{\@restonecolfalse\if@twocolumn\@restonecoltrue\onecolumn
     \else \newpage \fi \thispagestyle{empty}\c@page\z@
        \def\thefootnote{\fnsymbol{footnote}} }

\def\endtitlepage{\if@restonecol\twocolumn \else  \fi
        \def\thefootnote{\arabic{footnote}}
        \setcounter{footnote}{0}}  
\relax

\hybrid

\newfont{\Bbb}{msbm10 scaled 1\@ptsize00}
\newfont{\Bbbb}{msbm7 scaled 1\@ptsize00}

\newcommand{\DDD}{\raise-1pt\hbox{$\mbox{\Bbbb D}$}}

\newcommand{\UUU}{\raise-1pt\hbox{$\mbox{\Bbbb U}$}}

\newcommand{\z}{\raise-1pt\hbox{$\mbox{\Bbbb Z}$}}

\def\beq{\begin{equation}}
\def\eeq{\end{equation}}
\def\p{\partial}

\begin{document}

\begin{titlepage}

\title{Elliptic solutions of the semi-discrete BKP equation}

\author{D.~Rudneva\thanks{National Research University Higher School of Economics,
20 Myasnitskaya Ulitsa, Moscow 101000, Russian Federation;
Skolkovo Institute of Science and Technology, 143026 Moscow, Russian Federation; \newline
e-mail: missdaryarudneva@gmail.com
}
\and
 A.~Zabrodin\thanks{
Skolkovo Institute of Science and Technology, 143026 Moscow, Russian Federation;
ITEP NRC KI, 25 B.Cheremushkinskaya, Moscow 117218, Russian Federation;
e-mail: zabrodin@itep.ru}
}

\date{March 2020}
\maketitle

\vspace{-7cm} \centerline{ \hfill ITEP-TH-05/20}\vspace{7cm}

\begin{abstract}

We consider elliptic solutions of the semi-discrete BKP equation and
derive equations of motion for their poles.
The basic tool is the auxiliary linear problem for the wave function.

\end{abstract}

\end{titlepage}

\vspace{5mm}

\section{Introduction}

The dynamics of poles of singular solutions to nonlinear integrable
equations is a well known subject in mathematical physics 
\cite{AMM77,Krichever78,Krichever80,CC77}. 
In particular, it was shown
that the poles of singular solutions to the Kadomtsev-Petviashvili (KP) equation 
move as particles of the integrable Calogero-Moser many-body system
\cite{Calogero71,Calogero75,Moser75,OP81}. Rational, trigonometric and elliptic 
(doubly periodic in the complex plane) solutions
correspond respectively to rational, trigonometric or elliptic Calogero-Moser systems. 
In the most general elliptic case, the equations of motion are
\beq\label{int1}
\ddot x_i=4\sum_{k\neq i} \wp ' (x_i-x_k)
\eeq
($\wp$ is the Weierstrass $\wp$-function).

The method suggested by Krichever consists in substituting the pole ansatz not in the 
KP equation but in the auxiliary linear problem for it, which has the form 
of the non-stationary Schr\"odinger equation for the wave function $\psi$:
\beq\label{int2}
\p_t \psi =\p_x^2\psi +2U\psi , \quad U=-\sum_i \wp (x-x_i) +\mbox{const} .
\eeq 

Elliptic  solutions to the semi-discrete KP equation (and, more generally, to 
the matrix 2D Toda equation) were investigated in \cite{KZ95}. The corresponding linear problem
has the form of the differential-difference equation
\beq\label{int3}
\p_t \psi (x)=\psi (x+\eta )+u(x)\psi (x).
\eeq
In this case the dynamics of poles
is given by equations of motion of the integrable 
elliptic Ruijsenaars-Schneider system \cite{RS86}
(a relativistic version of the Calogero-Moser system):
\beq\label{int4}
\ddot x_i=\sum_{j\neq i}\dot x_i \dot x_j \frac{\wp '(x_i\! -\! x_j)}{\wp (\eta )-
\wp (x_i\! -\! x_j)}.
\eeq

Elliptic solutions 
to the B-version of the KP equation (BKP) 
\cite{DJKM83,DJKM82,DJKM82a,LW99,Tu07} 
were recently studied in
the paper \cite{RZ19}, see also \cite{Z1905}. It was shown that their poles move as particles
of a previously unknown many-body system with
equations of motion
\beq\label{int8}
\ddot x_i +6\sum_{j\neq i}(\dot x_i +\dot x_j)\wp '(x_i-x_j)-72\!\!
\sum_{j\neq k \neq i}\wp (x_i-x_j)\wp '(x_i-x_k)=0.
\eeq
The Hamiltonian structure and integrability of this system are to be further
investigated. 

In this paper we study elliptic solutions of the semi-discrete BKP equation
(with a discrete space variable and continuous time variable). 
We derive the equations
of motion for the dynamics of the poles (zeros of the tau-function).
Just as equations (\ref{int8}) are in some sense $B$-version of the Calogero-Moser
system, so the equations of motion in the semi-discrete case are a $B$-version 
of the Ruijsenaars-Schneider system. We also derive the commutation 
representation for the equations of motion which is a 
sort of the Manakov's triple representation
\cite{Manakov}. Like in the KP case, 
the main tool is the auxiliary linear problem for the wave function.

\section{The discrete BKP equation}

We begin with the continuous BKP hierarchy. Let ${\bf t}=\{ t_1, t_3, t_5, \ldots \}$
be an infinite set of continuous ``times'', the independent variables of the 
hierarchy. The dependent variable is the tau-function $\tau ({\bf t})$. The infinite
BKP hierarchy is encoded in the basic bilinear relation for the tau-function
\cite{DJKM82}
\beq\label{disc1}
\oint_{C_{\infty}}\frac{dz}{2\pi i z}\,
e^{\xi ({\bf t}, z)-\xi ({\bf t}', z)}
\tau \Bigl ({\bf t}-2[z^{-1}]\Bigr )\, \tau \Bigl ({\bf t}' +2[z^{-1}]\Bigr )=
\tau ({\bf t})\tau ({\bf t}')
\eeq
valid for any ${\bf t}$, ${\bf t}'$. Here we use the notation
$$
\xi ({\bf t}, z)=\sum_{k\geq 1, \, k \, {\rm odd}}t_k z^k, \quad
{\bf t}\pm 2[z^{-1}]=\left \{ t_1 \pm \frac{2}{z}, t_3 \pm \frac{2}{z^3},
t_5 \pm \frac{2}{z^5}, \ldots \right \}.
$$
The contour $C_{\infty}$ is a big circle around infinity.

The discrete BKP hierarchy is a subhierarchy of the continuous one.
The simplest discrete BKP equation is obtained as follows.
Put
\beq\label{disc2}
\tau (l,m,n)=\tau \Bigl ({\bf t}-2l[a^{-1}]-2m[b^{-1}]-2n[c^{-1}]\Bigr ),
\eeq
then setting $t_k'=t_k-2a^{-k}/k-2b^{-k}/k -2c^{-k}/k$ in
the bilinear relation (\ref{disc1}) one can calculate the integral 
using the residue calculus (it should be taken into account that the poles at
the points $a,b,c$ are outside the contour and the contour should be shrunk to
infinity). 
The result is that
$\tau (l,m,n)$ satisfies the discrete BKP equation \cite{Miwa82}
\beq\label{disc3}
\begin{array}{c}
(a+b)(a+c)(b-c)\tau (l+1)\tau (m+1, n+1)
\\ \\
\phantom{aaaaaaa}-(a+b)(b+c)(a-c)\tau (m+1)\tau (l+1, n+1)
\\ \\
\phantom{aaaaaaaaaaaaaa}+(a+c)(b+c)(a-b)\tau (n+1)\tau (l+1, m+1)
\\ \\
\phantom{aaaaaaaaaaaaaaaaaaaaa}=(a-b)(a-c)(b-c)\tau \, \tau (l+1, m+1, n+1).
\end{array}
\eeq
Here we explicitly write only those arguments that undergo shifts. 
Taking the limit $c\to \infty$, we get the semi-discrete BKP equation
\beq\label{disc4}
\begin{array}{c}
\displaystyle{
\tau \, \tau (l+1, m+1)\left (1+\frac{1}{a+b}\, 
\p_{t_1}\log \frac{\tau (l+1, m+1)}{\tau}\right )}
\\ \\
\displaystyle{
=\tau (l+1)\, \tau (m+1)\left (1+\frac{1}{a-b}\, 
\p_{t_1}\log \frac{\tau (l+1)}{\tau (m+1)}\right )}.
\end{array}
\eeq

The wave function $\psi (l,m;z)$ is introduced by the formula
\beq\label{disc5}
\psi (l,m;z)=e^{\xi ({\bf t},z)}\left (\frac{a-z}{a+z}\right )^{l}
\left (\frac{b-z}{b+z}\right )^{m}
\frac{
\tau \Bigl ({\bf t}-2l[a^{-1}]-2m[b^{-1}]-2[z^{-1}]
\Bigr )}{\tau \Bigl ({\bf t}-2l[a^{-1}]-2m[b^{-1}]\Bigr )}.
\eeq
It follows from (\ref{disc3}) that the wave function 
$\psi = \psi (l,m;z)$ satisfies the following
linear equation:
\beq\label{disc6}
\psi (m+1) -\psi (l+1)=\frac{a-b}{a+b}\, u (\psi (l+1, m+1)-\psi ),
\eeq
where
\beq\label{disc7}
u=\frac{\tau \, \tau (l+1, m+1)}{\tau (l+1)\tau (m+1)}.
\eeq
Tending $b\to \infty$, one obtains from (\ref{disc6}) the linear problem
\beq\label{disc8}
\p_{t_1}\Bigl (\psi (l)+\psi (l+1)\Bigr )=(v(l) +a)\Bigl (\psi (l)-\psi (l+1)\Bigr ),
\eeq
where
\beq\label{disc9}
v(l)=\p_{t_1}\log \frac{\tau (l+1)}{\tau (l)}.
\eeq

\section{Elliptic solutions}

Our aim is to study double-periodic (elliptic) in the variable $x=l\eta$ 
solutions of the semi-discrete BKP equation. Here $\eta$ is a parameter (a ``lattice spacing'').
For such solutions the tau-function is an 
``elliptic polynomial'' in the variable $x$:
\beq\label{es1}
\tau = C e^{Cx^2+Bxt_1}\prod_{i=1}^{N}\sigma (x-x_i)
\eeq
with some constants $C, B$, where 
$$
\sigma (x)=\sigma (x |\, \omega , \omega ')=
x\prod_{s\neq 0}\Bigl (1-\frac{x}{s}\Bigr )\, e^{\frac{x}{s}+\frac{x^2}{2s^2}},
\quad s=2\omega_1 m_1+2\omega_2 m_2 \quad \mbox{with integer $m_1, m_2$},
$$ 
is the Weierstrass 
$\sigma$-function with quasi-periods $2\omega_1$, $2\omega_2$ such that 
${\rm Im} (\omega_2/ \omega_1 )>0$. It is connected with the Weierstrass 
$\zeta$- and $\wp$-functions by the formulas $\zeta (x)=\sigma '(x)/\sigma (x)$,
$\wp (x)=-\zeta '(x)=-\p_x^2\log \sigma (x)$.
The roots $x_i=x_i(t_1)$ are assumed to be 
all distinct. 

We put $t_1=t$ and write the linear problem (\ref{disc8}) in the form
\beq\label{es2}
\p_{t}\Bigl (\psi (x)+\psi (x+\eta )\Bigr )=
(v(x) +\mu )\Bigl (\psi (x)-\psi (x+\eta )\Bigr ), \quad
v(x)=\p_t \log \frac{\tau (x+\eta )}{\tau (x)},
\eeq
where $\mu$ is a parameter (the former $a$ in (\ref{disc8})). For elliptic solutions
\beq\label{es3}
v(x)=B\eta +\sum_i \Bigl (\dot x_i \zeta (x-x_i)-\dot x_i \zeta (x+\eta -x_i)\Bigr ),
\eeq
where dot means the $t$-derivative. It is an elliptic function of $x$. 
Since the coefficient function $v$ is double-periodic, 
one can find double-Bloch solutions $\psi (x)$, i.e., solutions such that 
$\psi (x+2\omega_{\alpha} )=B_{\alpha}\psi (x)$ ($\alpha =1,2$)
with some Bloch multipliers $B_{\alpha}$.

The pole ansatz for the wave function is
\beq\label{es4}
\psi = e^{tz}\left (\frac{\mu -z}{\mu +z}\right )^{x/\eta}
\sum_{i=1}^N c_i \Phi (x-x_i, \lambda ),
\eeq
where the coefficients $c_i$ do not depend on $x$ (but do depend on $z$ and $t$).
Here the function $\Phi$ is defined as
$$
\Phi (x, \lambda )=\frac{\sigma (x+\lambda )}{\sigma (\lambda )\sigma (x)}\,
e^{-\zeta (\lambda )x}
$$
($\zeta$ is the Weierstrass $\zeta$-function). 
It has a simple pole
at $x=0$ with residue $1$:
$$
\Phi (x, \lambda )=\frac{1}{x}-\frac{1}{2}\, \wp (\lambda ) x +\ldots , \qquad 
x\to 0.
$$
The parameters $z$ and $\lambda$ 
are spectral parameters.
Using the quasiperiodicity properties of the function $\Phi$,
$$
\Phi (x+2\omega_{\alpha} , \lambda )=e^{2(\zeta (\omega_{\alpha} )\lambda - 
\zeta (\lambda )\omega_{\alpha} )}
\Phi (x, \lambda ),
$$
one can see that the wave function given by (\ref{es4}) 
is indeed a double-Bloch function with Bloch multipliers
$B_{\alpha}=e^{2(\omega_{\alpha} z + \zeta (\omega_{\alpha} )\lambda - 
\zeta (\lambda )\omega_{\alpha} )}$.
We will often suppress the second argument of $\Phi$ writing simply 
$\Phi (x)=\Phi (x, \lambda )$. 
We will also need the $x$-derivative 
$\Phi '(x, \lambda )=\p_x \Phi (x, \lambda )$. 

\section{Equations of motion for poles of elliptic solutions}

Let us substitute (\ref{es3}) and (\ref{es4}) into the linear problem (\ref{es2}).
The expression has obvious poles at $x=x_i$ and $x=x_i-\eta$. One should impose 
conditions on the coefficients $c_i$ which ensure cancellation of the poles. 
The second order poles cancel identically. From cancellation of the first order poles at
$x=x_i-\eta$ and $x=x_i$ we obtain the conditions
\beq\label{es5}
\left \{
\begin{array}{lll}
(z+\mu +B\eta )c_i +\dot c_i
&=& \displaystyle{\frac{z+\mu }{z-\mu }\, \dot x_i\sum_k
c_k \Phi (x_i-x_k -\eta )+\dot x_i\sum_{k\neq i}
c_k \Phi (x_i-x_k)}
\\ && \\
&&\displaystyle{-c_i \sum_k \dot x_k \zeta (x_i-x_k -\eta )+c_i
\sum_{k\neq i}\dot x_k \zeta (x_i-x_k)}
\\ && \\
(z-\mu -B\eta )c_i +\dot c_i
&=& \displaystyle{\frac{z-\mu }{z+\mu }\, \dot x_i\sum_k
c_k \Phi (x_i-x_k +\eta )+\dot x_i\sum_{k\neq i}
c_k \Phi (x_i-x_k)}
\\ && \\
&&\displaystyle{-c_i \sum_k \dot x_k \zeta (x_i-x_k +\eta )+c_i
\sum_{k\neq i}\dot x_k \zeta (x_i-x_k)}
\end{array}
\right.
\eeq
which should be valid for all $i=1, \ldots , N$.

Let us introduce the $N\! \times \! N$ 
matrices $A^{\pm}$, $A^0$ with matrix elements
$$
A^{\pm}_{ik}=\Phi (x_i-x_k \pm \eta ), \quad 
A^0_{ik}=(1-\delta_{ik})\Phi (x_i-x_k)
$$
and diagonal matrices $X, D^{\pm}, D^0$ with matrix elements
$X_{ik}=\delta_{ik}x_i$, 
$$
D^{\pm}_{ik}=\delta_{ik}\sum_j \dot x_j \zeta (x_i-x_j\pm \eta ), \quad
D^{0}_{ik}=\delta_{ik}\sum_{j\neq i} \dot x_j \zeta (x_i-x_j ), 
$$
then the conditions (\ref{es5}) can be written in the matrix form
as a system of linear equations for the column 
vector ${\bf c}=(c_1, \ldots , c_N)^T$:
\beq\label{es6}
\left \{ 
\begin{array}{l}
L{\bf c}=2\tilde \mu (z^2-\mu^2){\bf c}
\\ \\
\dot {\bf c}=M{\bf c},
\end{array}
\right.
\eeq
where $\tilde \mu =\mu +B\eta$ and the matrices $L$, $M$ have the form
\beq\label{es7}
L=(z+\mu )^2 \dot X A^- -(z-\mu )^2 \dot X A^+ +
(z^2-\mu^2)(D^+-D^-),
\eeq
\beq\label{es8}
M=-(z-\tilde \mu )I +\frac{z-\mu}{z+\mu}\, \dot X A^+ +\dot XA^0 -D^+ +D^0
\eeq
(here $I$ is the unity matrix). 

The system (\ref{es6}) is overdetermined. As a simple calculation shows, 
the compatibility condition is
\beq\label{es9}
(\dot L +[L,M]){\bf c}=0.
\eeq
We have:
$$
\begin{array}{lll}
\dot L +[L,M]&=&(z+\mu )^2\Bigl (\ddot X A^- +\dot X \dot A^-+
[\dot X A^-, \dot X A^0 -D^+ +D^0]\Bigr )
\\ && \\
&& -(z-\mu )^2\Bigl (\ddot X A^+ +\dot X \dot A^++
[\dot X A^+, \dot X A^0 -D^- +D^0]\Bigr )
\\ && \\
&& +(z^2-\mu^2)\Bigl (\dot D^+-\dot D^- -[\dot XA^+, \dot XA^-]+
[D^+-D^-, \dot X A^0]\Bigr ).
\end{array}
$$
With the help of the identities
\beq\label{es10}
\Phi (x, \lambda )\Phi (y, \lambda )=\Phi (x+y, \lambda )\Bigl (
\zeta (x)+\zeta (y)-\zeta (x+y+\lambda )+\zeta (\lambda )\Bigr ),
\eeq
\beq\label{es11}
\Phi (x, \lambda )\Phi (-x, \lambda )=\wp (\lambda )-\wp (x)
\eeq
one can see, by a straightforward calculation, that
$$
\dot X \dot A^{\pm}+
[\dot X A^{\pm}, \dot X A^0 -D^{\mp} +D^0]=
(D^++D^--2D^0)\dot XA^{\pm},
$$
$$
\dot D^+-\dot D^- -[\dot XA^+, \dot XA^-]+
[D^+-D^-, \dot X A^0]=W^+-W^-,
$$
where $W^{\pm}$ are diagonal matrices with matrix elements
$$
W^{\pm}_{ii}=\sum_j \ddot x_j \zeta (x_i-x_j \pm \eta )+\sum_j
\dot x_j^2 \wp (x_i-x_j\pm \eta ).
$$
Therefore, we have the matrix identity
\beq\label{es12}
\dot L +[L,M]=R\Bigl (L-2\tilde \mu (z^2-\mu^2)I\Bigr )+(z^2-\mu^2)P,
\eeq
where $R,P$ are the diagonal matrices
\beq\label{es13}
R=\ddot X \dot X^{-1}+D^++D^--2D^0,
\eeq
\beq\label{es14}
P=W^+-W^--R(D^+-D^--2\tilde \mu I).
\eeq
We see that the compatibility condition (\ref{es9}) means that all elements
of the diagonal matrix $P$ should be equal to zero ($P_{ii}=0$ for all $i$).
This yields the equations of motion for the $x_i$'s. 

Using the standard identities for the Weierstrass functions, one can bring the
equations of motion to the form
\beq\label{es15}
\begin{array}{l}
\displaystyle{
\sum_{j\neq i}(\ddot x_i \dot x_j-\dot x_i\ddot x_j)\left (
\frac{\wp '(\eta)}{\wp (x_{ij})-\wp (\eta)}-2\zeta(\eta)\right )}
\\ \\
\displaystyle{\phantom{aaaaaaaaa}
+\sum_k \sum_{j\neq i,k}\dot x_i\dot x_j\dot x_k 
\frac{\wp '(x_{ij})}{\wp (x_{ij})-\wp (\eta)}
\left (\frac{\wp '(\eta)}{\wp (x_{ik})-\wp (\eta)}-2\zeta(\eta)\right )
}
\\ \\
\displaystyle{\phantom{aaaaaaaaaaaaaaaaa}
+2\tilde \mu \ddot x_i +2\tilde \mu \sum_{j\neq i}
\dot x_i\dot x_j \frac{\wp '(x_{ij})}{\wp (x_{ij})-\wp (\eta)}=0,}
\end{array}
\eeq
where $x_{ij}=x_i-x_j$. In contrast to the equations of motion
(\ref{int4}), (\ref{int8}), this system of linear equations 
is not resolved with respect to the $\ddot x_i$'s. 
In the rational limit ($\wp (x)\to 1/x^2$) the equations of motion
for poles of rational solutions are
\beq\label{es16}
\sum_{j\neq i}\frac{\ddot x_i\dot x_j-\dot x_i\ddot x_j}{x_{ij}^2-\eta^2}
+\sum_k \! \sum_{j\neq i,k}\frac{2\eta^2 \dot x_i\dot x_j\dot x_k}{x_{ij}(x_{ij}^2-\eta^2)
(x_{ik}^2-\eta^2)}
+\frac{\tilde \mu}{\eta}\ddot x_i +\tilde \mu\sum_{j\neq i}
\frac{2\eta \dot x_i\dot x_j}{x_{ij}(x_{ij}^2-\eta^2)}=0.
\eeq

The commutation representation of the equations of motion follows from
(\ref{es12}). It is
\beq\label{es17}
\dot L +[L,M]=R\Bigl (L-2\tilde \mu (z^2-\mu^2)I\Bigr )
\eeq
which is a sort of the Manakov's triple representation \cite{Manakov}. 

\section{Concluding remarks}

The equations of motion (\ref{es15}) for poles of elliptic solutions to the
semi-discrete BKP equation together with their 
Manakov's triple representation (\ref{es17}) are the main results of the paper. The resulting
system of equations of motion is not resolved with respect to accelerations 
$\ddot x_i$. 
Such a non-resolved form of equations of motion was previously known 
for elliptic solutions 
to the Novikov-Veselov equation \cite{Z1905}. However, in this case the system
admits an explicit solution in the rational limit. We do not know whether the system
(\ref{es16}) admits an explicit solution for $\ddot x_i$. It is not also clear how
to construct integrals of motion. These are problems for future investigation.

\section*{Acknowledgments}

The work of A.Z. was supported by the 
Russian Science Foundation under grant 19-11-00275.


\begin{thebibliography}{99}

\bibitem{AMM77}
H. Airault, H.P. McKean, and J. Moser, {\it Rational and 
elliptic solutions of the
Korteweg-De Vries equation and a related many-body problem},
Commun. Pure Appl. Math., {\bf 30} (1977) 95-148.



\bibitem{Krichever78}
I.M. Krichever, {\it Rational solutions of the Kadomtsev-Petviashvili
equation and integrable systems of $N$ particles on a line},
Funct. Anal. Appl. {\bf 12:1} (1978) 59-61.


\bibitem{Krichever80} I.M. Krichever, {\it Elliptic solutions of the Kadomtsev-Petviashvili
equation and integrable systems of particles}, Funk. Anal. i Ego Pril. {\bf 14:4} (1980) 45-54
(in Russian); English translation:
Functional Analysis and Its Applications {\bf 14:4} (1980) 282–-290.

\bibitem{CC77} D.V. Chudnovsky, G.V. Chudnovsky, {\it Pole expansions of non-linear
partial differential equations}, Nuovo Cimento {\bf 40B} (1977) 339-350.

\bibitem{Calogero71}
F. Calogero, {\it Solution of the one-dimensional
$N$-body problems with quadratic
and/or inversely quadratic pair potentials}, J. Math. Phys.
{\bf 12} (1971) 419-436.

\bibitem{Calogero75} F. Calogero, {\it Exactly solvable one-dimensional many-body
systems}, Lett. Nuovo Cimento {\bf 13} (1975) 411-415.

\bibitem{Moser75}
J. Moser, {\it Three integrable Hamiltonian systems connected with isospectral
deformations}, Adv. Math. {\bf 16} (1975) 197-220.

\bibitem{OP81} M.A. Olshanetsky and A.M. Perelomov, {\it Classical integrable
finite-dimensional systems related to Lie algebras}, Phys. Rep. {\bf 71} (1981) 313-400.

 \bibitem{KZ95} I. Krichever and A. Zabrodin, {\it 
Spin generalization of the Ruijsenaars-Schneider model, non-abelian 2D
Toda chain and representations of Sklyanin algebra}, Uspekhi Mat. Nauk
{\bf 50} (1995) 3-56 (in Russian) (English translation: 
Russ. Math. Surv., {\bf 50} (1995) 1101-1150).

\bibitem{RS86} S.N.M. Ruijsenaars and H. Schneider, {\it 
A new class of integrable systems and its relation to
solitons}, Annals of Physics {\bf 146} (1986) 1--34.
 

\bibitem{DJKM83} E. Date, M. Jimbo, M. Kashiwara and T. Miwa,
{\it Transformation groups for soliton equations: Nonlinear integrable systems --
classical theory and quantum theory} (Kyoto, 1981). Singapore: World Scientific,
1983, 39-119.

\bibitem{DJKM82} E. Date, M. Jimbo, M. Kashiwara and T. Miwa,
{\it Transformation groups for soliton equations IV. A new hierarchy
of soliton equations of KP type}, Physica D {\bf 4D} (1982) 343-365.

\bibitem{DJKM82a} E. Date, M. Jimbo, M. Kashiwara and T. Miwa,
{\it Quasi-periodic solutions of the orthogonal KP equation.
Transformation groups for soliton equations V}, Publ. RIMS, Kyoto Univ.
{\bf 18} (1982) 1111-1119.

\bibitem{LW99} I. Loris and R. Willox, {\it Symmetry reductions of the BKP
hierarchy}, Journal of Mathematical Physics {\bf 40} (1999) 1420-1431.

\bibitem{Tu07} M.-H. Tu, {\it On the BKP Hierarchy: Additional Symmetries,
Fay Identity and Adler–-Shiota–-van Moerbeke Formula}, Letters in Mathematical Physics
{\bf 81} (2007) 93-105.

\bibitem{RZ19} D. Rudneva and A. Zabrodin, {\it Dynamics of poles of elliptic
solutions to BKP equation}, arXiv:1903.00968, to be published in 
Journal of Physics A.

\bibitem{Z1905} A. Zabrodin, {\it Elliptic solutions to integrable nonlinear
equations and many-body systems}, Journal of Geometry and Physics,
{\bf 146} (2019) 103506, arXiv:1905.11383.

\bibitem{Manakov} S. Manakov, {\it Method of inverse scattering problem
and two-dimensional evolution equations}, Uspekhi Mat. Nauk {\bf 31} (1976)
245-246.

\bibitem{Miwa82} T. Miwa, {\it On Hirota's difference equations},
Proc. Japan Acad. {\bf 58} Ser. A (1982) 9-12.

\end{thebibliography}
\end{document}